\documentclass[
 reprint,
 superscriptaddress,
 showpacs,preprintnumbers,
 amsmath,amssymb,
 aps,
 pra,
 longbibliography,
lengthcheck,
]{revtex4-1}

\usepackage{makecell}% Table
\usepackage{graphicx}% Include figure files
\usepackage{dcolumn}% Align table columns on decimal point
\usepackage{bm}% bold math
\usepackage{color}
\usepackage{ulem}
\usepackage{hyperref}% add hypertext capabilities
\usepackage{float}
\usepackage{braket}

\begin{document}

\preprint{APS/123-QED}

\title{
  Controlling resonant spin photocurrent using magnetic field:\\
  Application to a magnetoelectric oxide Cr$_2$O$_3$
}

\author{Zhuo-Cheng Gu}
\affiliation{
  Department of Physics, Institute of Science Tokyo, Meguro, Tokyo, 152-8551, Japan
}

\author{Hiroaki Ishizuka}
\affiliation{
  Department of Physics, Institute of Science Tokyo, Meguro, Tokyo, 152-8551, Japan
}

\date{\today}

\begin{abstract}
  We studied the magnon spin photocurrent in effective spin models for Cr$_2$O$_3$, a material known for the magnetoelectric (ME) effect.
  Using the nonlinear response theory, we show that the magnon spin current can be generated by both linearly and circularly polarized electromagnetic waves via one-magnon processes.
  For linearly polarized waves, the spin current arises even in the absence of a static magnetic field.
  In contrast, circularly polarized waves induce a spin current only when a static magnetic field is present, and the current reverses its direction upon inversion of the field.
  The sensitivity to the external magnetic field and the contrasting behaviors to the linearly and circularly polarized electromagnetic waves delineate the spin photocurrent from other competing contributions, such as the spin pump and inhomogeneous heating, facilitating experimental verification.
  Our results show that Cr$_2$O$_3$ is an interesting candidate for the experimental investigation of the magnon photocurrent.
\end{abstract}

\pacs{
}% PACS, the Physics and Astronomy
% Classification Scheme.

\maketitle

%%%%   Introduction   %%%%

\section{Introduction}
The interplay between light and magnetism gives rise to rich phenomena, which allow both the detection and control of magnetic states~\cite{Kirilyuk2010a,Nemec2018a,Siegrist2019a,Afanasiev2021a,Walowski2016a}.
In metals, magnetism can be manipulated through itinerant electrons via the spin transfer effect~\cite{Berger1984a,Maekawa2012a}, while spin currents can be generated through spin pumping~\cite{Kajiwara2010a,Heinrich2011a,Ohnuma2014a}.
In insulating materials, light has been proven to be a powerful tool for controlling magnetism, including laser-induced magnetic switching~\cite{Kimel2005a,Stanciu2007a,Kirilyuk2010a,Mukai2016a} and modulation of spin-wave propagation with focused light~\cite{Satoh2012a,Sato2017a}.
Moreover, it has been shown that light can control exchange interactions~\cite{Mentink2015a,Ono2017a} and manipulate magnetic textures~\cite{Mochizuki2010a,Koshibae2014a,Sato2016a}.
The versatility of light in controlling magnetism makes it attractive for both fundamental research and technological applications.

In view of controlling magnetic excitations, nonlinear responses have garnered interest.
One notable example is the magnon analog of the bulk photovoltaic effect~\cite{vBaltz1981a,Sturman1992a,Sipe2000a,Tan2016a,Tokura2018a} --- a dc electric current induced by light in a bulk material, similar to solar cells --- was predicted to occur by optically exciting magnons~\cite{Proskurin2018a,Ishizuka2019b} and spinons~\cite{Ishizuka2019a}.
Phenomenologically, the spin photocurrent is given by $J_s = \sigma^{(2)} h(\omega) h(-\omega)$,
where $J_s$ denotes the spin current and $h(\omega)$ is the ac magnetic field of frequency $\omega$.
Similar to the electronic photogalvanic effect, this spin current is expected to emerge in noncentrosymmetric mateirals, such as in two-dimensional magnets~\cite{Ishizuka2022a} and multiferroic oxides~\cite{Fujiwara2023a}.
This phenomenon holds promise for optospintronics, providing another method to control magnetic materials and spin current.
However, experimental verification remains challenging due to competing effects such as spin pumping and spin currents generated by inhomogeneous heating.

In this work, we study Cr$_2$O$_3$~\cite{Brockhouse1953a,Shi2009a,Skovhus2022a,Ghotekar2021a,Lebreau2014a,Zhu2023a,Shiratsuchi2021a}, a material known for its electromagnetic properties, as a candidate to realize the magnon spin photocurrent by the single-magnon process~\cite{Ishizuka2022a}. Focusing on four-sublattice spin models in one~\cite{Izuyama1963a} and three dimensions~\cite{Samuelsen1970a}, both of which serve as the effective spin model for Cr$_2$O$_3$, we theoretically investigate the nonlinear spin current induced by linearly and circularly polarized electromagnetic waves using nonlinear response theory.
In the one-dimensional model, we elucidate the fundamental properties of the spin photocurrent in Cr$_2$O$_3$.
We show that linearly polarized waves generate a spin current even in the absence of an external static magnetic field. In contrast, circularly polarized waves require a finite static magnetic field to induce a spin current, with the current direction reversing upon inversion of the magnetic field.
These distinct behaviors, which differ from conventional spin pumping and inhomogeneous heating effects, provide a clear signature for distinguishing the spin photocurrent from competing phenomena.

This paper is organized as follows. In Sec.~\ref{sec:model}, we introduce the model and method we used to study the magnon spin photocurrent.
In Sec.~\ref{sec:1D}, we consider a one-dimensional model for Cr$_2$O$_3$, which we use as a toy model to understand the fundamental behavior of the magnon spin photocurrent.
The magnon spin photocurrent in a three-dimensional model for Cr$_2$O$_3$ is studied in Sec.~\ref{sec:3D}. Section~\ref{sec:discussion} is devoted to the conclusion and discussion.

\section{Model and Method}\label{sec:model}

\subsection{Collinear magnet}

\begin{figure}[t]
  \centering
  \includegraphics[width=\linewidth]{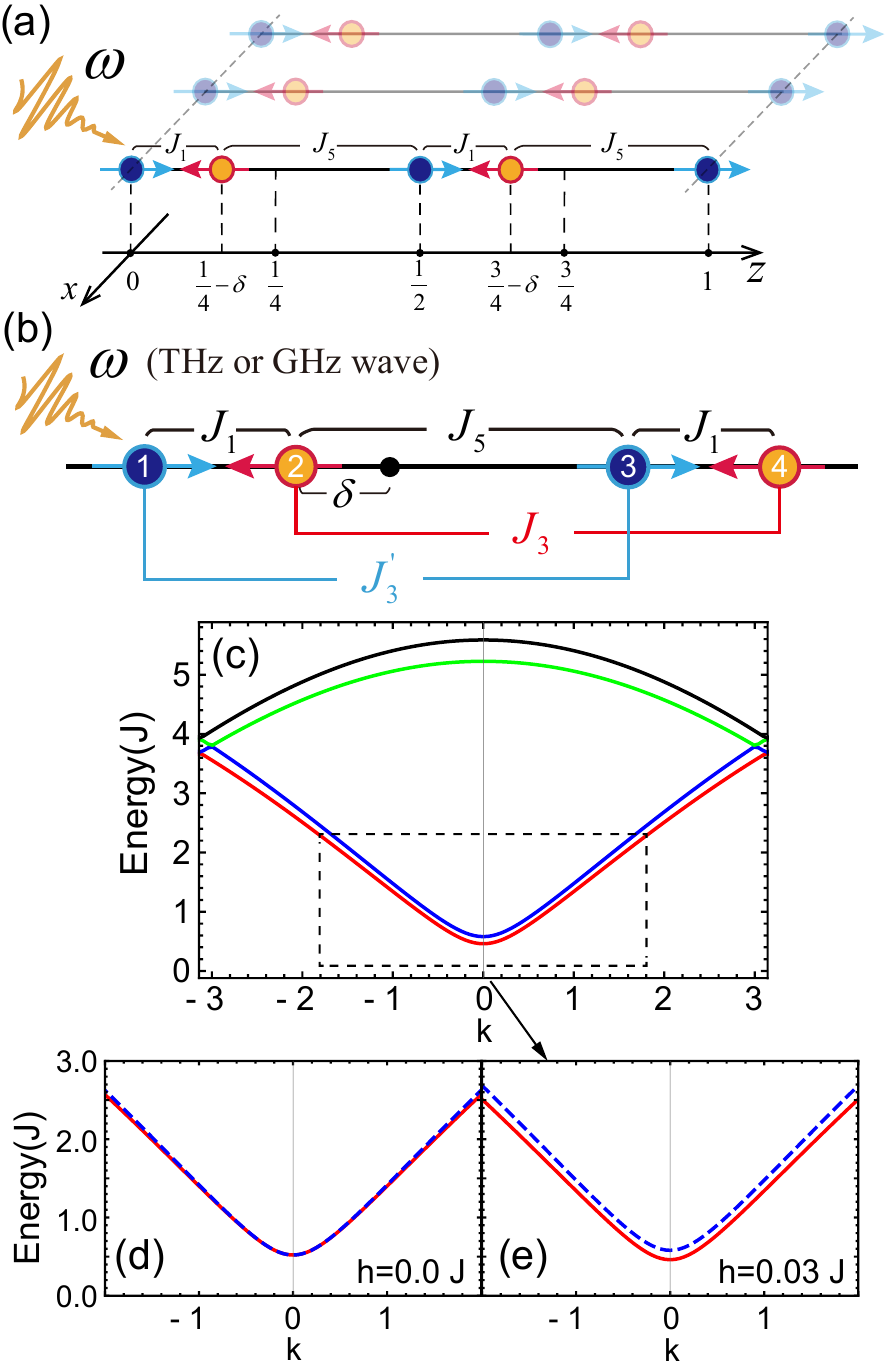}
  \caption{
    Schematic of the spin model and its magnon dispersion. (a) A noncentrosymmetric quasi-one-dimensional magnetic system composed of weakly coupled spin chains with nearest-neighbor interactions, $J_1$ and $J_5$, and displacement $\delta$.
    (b) A schematic of the one-dimensional model studied in Sec.~\ref{sec:1D}. The numbers $1$-$4$ denote the sublattice.
    (c) Magnon dispersion of the one-dimensional model with $h=0.03J$ and enlarged views of the lower magnon band near the $\Gamma$ point with (d) $h = 0$ and (e) $h = 0.03~J$.
    The results are for $J_1 = -1$, $J_5 = -0.5$, $J_3 = 0.1$, $J_3' = 0.08$, $D_z = 0.005$, and $\delta = 0.1$.
  }
  \label{fig:method_model}
\end{figure}

In this work, we consider spin models with uniaxial anisotropy whose Hamiltonian reads
\begin{align}
  \mathcal{H} = & - \sum_{\langle jp, iq \rangle} J_{jp,iq}\bm S_{jp} \cdot \bm S_{iq} - D_z \sum_{jp} \left(S_{jp}^z\right)^2 - h \sum_{jp} S_{jp}^z.\label{eq:H0}
\end{align}
Here, $\bm S_{ju} = (S_{ju}^x, S_{ju}^y, S_{ju}^z)$ is the spin on the sublattice $u$ in the $j$th unit cell, $J_{jp,iq}$ is the Heisenberg exchange interaction between the two spins, $\bm S_{jp}$ and $\bm S_{iq}$, and $D_z\ge0$ is the Ising anisotropy.
The last term is the Zeeman coupling between the spins and the magnetic field, where $h = g \mu_B H$ denotes the Zeeman energy with $g \approx 2$ being the Land\'e g-factor, $\mu_B$ is the Bohr magneton, and $H$ is the external magnetic field.
An example of the model is shown in Fig.~\ref{fig:method_model}(a), which we will consider in Sec.~\ref{sec:1D}.

As an interaction between light and the spins, we consider the Zeeman coupling between the spins and the ac magnetic field.
The Hamiltonian reads
\begin{align}\label{eq:general_perturbation_defination}
  \mathcal{H}'(t) = - B_x(t) \mu_B g \sum_{j,p} S_{j,p}^x - B_y (t) \mu_B g \sum_{j,p} S_{j,p}^y,
\end{align}
where $t$ is the time and $B_\alpha(t)$ ($\alpha = x,y$) is the ac magnetic field along the $\alpha$ axis.

Within the linear spin-wave approximation, the spin Hamiltonian reduces to a free boson Hamiltonian through
the Holstein-Primakoff transformation~\cite{Holstein1940a}.
The resulting Hamiltonian reads
\begin{align}
  \mathcal{H} = \frac12\sum_{\bm k}\psi_{\bm k}^\dagger \hat{H}_{\bm k} \psi_{\bm k},
\end{align}
where $\psi_{\bm k}=\left(b_{1\bm k},\dots,b_{n_{uc}\bm k},b_{1,-\bm k}^\dagger,\dots,b_{n_{uc},-\bm k}^\dagger\right)^T$ is an array of boson annihilation (creation) operators $b_{i\bm k}$ ($b_{i\bm k}^\dagger$) with $\bm k$ being the momentum and $i$ denotes the sublattice.
This Hamiltonian is readily diagonalized by a well-known protocol~\cite{Colpa1978a}; the Hamiltonian becomes
\begin{align}
  \mathcal{H} = \frac12\sum_{\bm k}\phi_{\bm k}^\dagger \hat{E}_{\bm k} \phi_{\bm k},
\end{align}
where $\hat{E}_{\bm k}$ is a diagonalized matrix and $\phi_{\bm k} ={(\alpha_{j,\bm k}, \alpha_{j,-\bm k}^\dagger)}^T$.
In the below, we mainly study the spin current within this linear spin wave theory.

\subsection{Magnon spin photocurrent}

The spin current operator for $S^z$ is defined using the continuity equation.
As the spin angular momentum for $S^z$ is conserved in the Hamiltonian in Eq.~\eqref{eq:H0}, the density of $S^z$ at position $\bm r$, $S^z(\bm r)$, obeys the continuity equation $\nabla \cdot \bm{\mathcal J} = - \partial_t S^z = i[S^z,H]$.
By deriving the Heisenberg equation of motion for $S^z$ and comparing it with the continuity equation, the spin current operator is defined as
\begin{align}
  \mathcal J^{\mu} = \frac{1}{2N} \sum_{\langle jp, iq \rangle} J_{jp,iq} \left[(\bm{r_{jp}})_\mu - (\bm{r_{iq}})_\mu \right](S_{jp}^x S_{iq}^y - S_{jp}^y S_{iq}^x),\label{eq:Jspin-gen}
\end{align}
where ${\mathcal J}^\mu$ is the spin current flowing along the $\mu$ axis.

To study the nonlinear spin current of Eq.~\eqref{eq:H0},
we calculate the nonlinear spin current conductivity by combining the linear spin wave approximation~\cite{
  Holstein1940a} and the nonlinear response theory~\cite{Ishizuka2019b,Ishizuka2022a}. The nonlinear conductivity $\sigma_{\mu \nu \lambda}^{(2)}(\Omega;\omega,-\omega)$ is defined by
\begin{align}\label{eq:general_spincurrent_defination}
  J_\mu (\Omega) = \int \sum_{\nu \lambda} \sigma_{\mu \nu \lambda}^{(2)}(\Omega;\omega,-\omega) B_{\nu}(\omega) B_{\lambda}(\Omega-\omega) d\omega,
\end{align}
where $J_\mu(\Omega) = \int \langle \mathcal J_{\mu} \rangle e^{-i\Omega}dt$ with $\langle O\rangle$ being the thermal average of $O$, and $B_\alpha(\omega) = \int B_\alpha(t) e^{-i\omega}dt$. For the external field in Eq.~\eqref{eq:general_perturbation_defination}, the nonlinear conductivity is given by~\cite{Ishizuka2022a},
\begin{align}
  \label{eq:method_general_conductivity_defination_energy}
   & \sigma_{\mu\nu\lambda}^{(2)}(0;\omega,-\omega) = -\frac{1}{4\pi V} \sum_{l = 1}^{2N} \sum_{m = 1}^{2N}\frac{\gamma_l \gamma_m}{\hbar\omega-\gamma_l E_{l \bm 0}-i\eta} \nonumber                                                                                      \\
   & \times \frac{[\tilde{\beta}_{\bm 0}^\nu]_l [\tilde{\beta}_{\bm 0}^\lambda]_{m+N\gamma_m} \left([\tilde{\mathcal{J}}_{{\mu}, \bm 0}]_{l,m}+[\tilde{\mathcal{J}}_{{\mu}, \bm 0}]_{m+N\gamma_m,l+N\gamma_l}\right)}{-\gamma_l E_{l \bm 0} +\gamma_m E_{m \bm 0} -i\eta},
\end{align}
where $\hbar$ is the reduced Planck constant, $V$ is the volume of the system, and $N$ is the number of sublattices, and $\gamma_l$ reads
\begin{align}
  \gamma_l= & \left\{\begin{array}{rl}
                       1  & (1 \leq l \leq N)     \\
                       -1 & (N+1 \leq l \leq 2N).
                     \end{array}\right.
\end{align}
Here, $[\tilde{\mathcal{J}}_{\mu, \bm 0}]_{lm} = \langle l |\tilde{\mathcal{J}}_{\mu,\bm 0} | m \rangle$ is the matrix elements for $\tilde{\mathcal{J}}_{\mu,\bm 0}$ where $|m \rangle$ is the ket vector of $m$th magnon state with $\bm k = \bm 0$, $E_{m \bm 0}$ is the eigenenergy of $|m \rangle$, and $\eta$ is the inverse of the magnon relaxation time.
The $[\tilde{\beta}_{\bm 0}^\nu]_{l}$ is the coefficient of ${\mathcal H}'$ in the eigenstate basis defined by
\begin{align}\label{eq:model_perturbation}
  \mathcal{H}'(t) = - \sum_{\substack{l \\\nu=x,y,z}} B_\nu(t)\left([\tilde{\beta}_{\bm 0}^\nu]_{l}b_{\bm 0l}+[\tilde{\beta}_{\bm 0}^\nu]_{l+N}b_{\bm 0l}^\dagger\right),
\end{align}
with $[\tilde{\beta}_{\bm 0}^\nu]_{i+N} = ([\tilde{\beta}_{\bm 0}^\nu]_{i})^*$.

For the case of linearly polarized light, and with $\omega_{n,\bm k}$ being an even function of $\bm k$, Eq.~\eqref{eq:method_general_conductivity_defination_energy} reads (See Appendix~\ref{sec:Berry} for the derivation),
\begin{align}
   & \sigma_{\mu\nu\nu}^{(2)}(0;\omega,-\omega) =-\frac{i}{2\pi V} \nonumber                                                                                                                                                                                 \\
   & \times \sum_{l} \frac{1}{\hbar\omega - \gamma_l \hbar\omega_{l\bm{0}} - i\eta} \tilde{n}_l \left[ \partial_{k_\lambda}\phi_{l \bm 0} + \gamma_l \mathcal{A}_{ll}^\mu(\bm 0) \right] \left|\tilde{\beta}_{l \bm 0}^\mu\right|^2,\label{eq:shift-current}
\end{align}
similar to the formula for phonon Peltier effect~\cite{Ishizuka2024a,Ishizuka2025a}.
Here, $\mathcal{A}_{lm}^\mu(\bm k)=i\langle u_{l\bm  k}|\tau^z|\partial_{k_\mu}u_{m\bm k}\rangle$, is the Berry connection~\cite{Cheng2016a, Zyuzin2016a} of antiferromagnetic magnons, which is a generalization of the ones for ferromagnets~\cite{Matsumoto2011a}, and $\phi_{l\bm k}^\mu$ is the phase of $\tilde\beta_{l\bm k}^\mu$, i.e. $\tilde\beta_{l\bm k}^\mu=|\tilde\beta_{l\bm k}^\mu|e^{i\phi_{l\bm k}^\mu}$.
The formula resembles that of the shift current~\cite{vBaltz1981a,Sipe2000a}, which is related to the Berry connection of the electrons.

\section{One-dimensional spin chain}\label{sec:1D}

First, we consider a simplified one-dimensional (1D) spin chain, which was introduced as an effective model for Cr$_2$O$_3$~\cite{Izuyama1963a}; the schematic of the spin chain is shown in Fig.~\ref{fig:method_model}(a).
The Hamiltonian is given by
\begin{align}
  \label{eq:1D_Hamiotlnian_initial}
   & \mathcal{H}^{1D} = - \sum_{\langle ip, jq \rangle}\left(\mathcal{H}_{NN}^{1D} + \mathcal{H}_{NNN}^{1D}\right) + \mathcal{H}_D + \mathcal{H}_h \nonumber \\
   & \mathcal{H}_{NN}^{1D} = {J_1}{{\bm S}_{j,1}} \cdot {{\bm S}_{j,2}} + {J_5}{{\bm S}_{j,2}} \cdot {{\bm S}_{j,3}} \nonumber                               \\
   & \qquad \qquad + {J_1}{{\bm S}_{j,3}} \cdot {{\bm S}_{j,4}} + {J_5}{{\bm S}_{j,4}} \cdot {{\bm S}_{j + 1,1}}, \nonumber                                  \\
   & \mathcal{H}_{NNN}^{1D} =  {J_3}{{\bm S}_{j,1}} \cdot {{\bm S}_{j,3}} + {{J'}_3}{{\bm S}_{j,2}} \cdot {{\bm S}_{j,4}} \nonumber                          \\
   & \qquad\qquad + {J_1}{{\bm S}_{j,3}} \cdot {{\bm S}_{j + 1,1}} + {{J'}_3}{{\bm S}_{j,4}} \cdot {{\bm S}_{j + 1,2}} \nonumber                             \\
   & \mathcal{H}_D = - {D_z}{\sum\limits_{j,p} {(S_{j,p}^z)} ^2} , \nonumber                                                                                 \\
   & \mathcal{H}_h = - h\sum_{j,p} S_{j,p}^z,
\end{align}
where $\mathcal{H}_{NN}^{1D}$ and $\mathcal{H}_{NNN}^{1D}$ are the nearest neighbor (NN) interaction and the next nearest neighbor (NNN) interaction, respectively, $H'$ is the Zeeman coupling between the spins and the external electromagnetic field, and $\mathcal{H}_D$ denotes the uniform anisotropy. In the below, we assume $J_3,\;J_3' >0$ and $J_1,\;J_5 < 0$.

Within the linear spin wave theory, the magnon Hamiltonian for $H^{1D}$ becomes
\begin{widetext}
  \begin{align}\label{eq:1D_Hamiltonian_final_form}
    \mathcal{H}^{1D} & = - \left(J_1 + J_5 - 2D_z \right) S \sum_k (a_{k}^\dagger a_{k} + b_{-k}^\dagger b_{-k} + c_{k}^\dagger c_{k} + d_{-k}^\dagger d_{-k}) \nonumber                                                                                                                                      \\
                     & \quad + 2J_3S\sum_k \left[(a_{k}^\dagger a_{k} + c_{k}^\dagger c_{k}) \right] + 2J_3'S\sum_k \left[(b_{-k}^\dagger b_{-k} + d_{-k}^\dagger d_{-k})\right] + h \sum_k (a_k^\dagger a_k - b_k^\dagger b_k + c_k^\dagger c_k - d_k^\dagger d_k), \nonumber                                \\
                     & \quad - J_1 S \sum_k \left(e^{-ik(\frac{1}{4}-\delta)} a_{k} b_{-k} + e^{ik(\frac{1}{4} - \delta)} a_{k}^\dagger b_{-k}^\dagger \right) - J_5 S \sum_k \left(e^{ik(\frac{1}{4} + \delta)} b_{-k} c_{k} + e^{-ik(\frac{1}{4} + \delta)} b_{-k}^\dagger c_{k}^\dagger \right) \nonumber  \\
                     & \quad - J_1 S \sum_k \left(e^{-ik(\frac{1}{4} - \delta)} c_{k} d_{-k} + e^{ik(\frac{1}{4} - \delta)} c_{k}^\dagger d_{-k}^\dagger \right) - J_5 S \sum_k \left(e^{ik(\frac{1}{4} + \delta)} d_{-k} a_{k} + e^{-ik(\frac{1}{4} + \delta)}d_{-k}^\dagger a_{k}^\dagger \right) \nonumber \\
                     & \quad + 2J_3S\sum_k \left[ - \cos \left(\frac{k}{2}\right) a_{k}^\dagger c_{k} - \cos \left(\frac{k}{2}\right) a_{k} c_{k}^\dagger \right] + 2J_3'S\sum_k \left[-\cos\left(\frac{k}{2}\right) b_{-k}^\dagger d_{-k} - \cos\left(\frac{k}{2}\right) b_{-k} d_{-k}^\dagger \right],
  \end{align}
  where we have neglected the constant terms.
  Here, $a_{k}$ ($a_{k}^\dagger$), $b_{k}$ ($b_{k}^\dagger$), $c_{k}$ ($c_{k}^\dagger$), and $d_{k}$ ($d_{k}^\dagger$) are the annihilation (creation) operators for the magnons on sublattices 1-4, respectively.
  Similarly, the spin current operator reads\begin{align}\label{eq:1D_spincurrent_final_form}
    \mathcal J^{z} & = \frac{iS}{{N}} \sum_{k} \left\{\left(\frac{1}{4}-\delta \right)J_1 \left[e^{ik(\frac{1}{4}-\delta)} b_{-k}^\dagger a_k^\dagger - e^{-ik(\frac{1}{4}-\delta)}b_{-k}a_{k} \right] + \left(\frac{1}{4}+\delta\right) J_5 \left[ e^{ik(\frac{1}{4}+\delta)}d_{-k} a_k - e^{-ik(\frac{1}{4}+\delta)}d_{-k}^\dagger a_k^\dagger \right] \right\} \nonumber  \\
                   & + \frac{iS}{{N}}\sum_{k} \left\{\left(\frac{1}{4}-\delta \right)J_1 \left[ e^{ik(\frac{1}{4}-\delta)} d_{-k}^\dagger c_k^\dagger - e^{-ik(\frac{1}{4}-\delta)} d_{-k} c_k \right] + \left(\frac{1}{4}+\delta \right) J_5 \left[e^{ik(\frac{1}{4}+\delta)}b_{-k} c_k - e^{-ik(\frac{1}{4}+\delta)} b_{-k}^\dagger c_k^\dagger \right] \right\} \nonumber \\
                   & - \frac{S}{{N}} \sum_k \sin \left(\frac{k}{2} \right) \left[J_3(c_k a_k^\dagger + c_k^\dagger a_k) - J'_3 (d_{-k}^\dagger b_{-k} + d_{-k} b_{-k}^\dagger) \right].
  \end{align}
\end{widetext}
Note that, the $k=0$ part of ${\mathcal J}^z$ does not depend on $J_3$ as $\sin(k/2) = 0$, which will be relevant to the discussion below.

The dispersion of the magnon band is shown in Figs.~\ref{fig:method_model}(c) - \ref{fig:method_model}(e).
All magnon bands are doubly degenerate when $h=0$ as shown in Fig.~\ref{fig:method_model}(d).
With a non-zero $h$, the degeneracy is lifted due to the Zeeman splitting and results in four distinct energy bands [Figs.~\ref{fig:method_model}(c) \& \ref{fig:method_model}(e)].

\begin{figure}
  \centering
  \includegraphics[width=\linewidth]{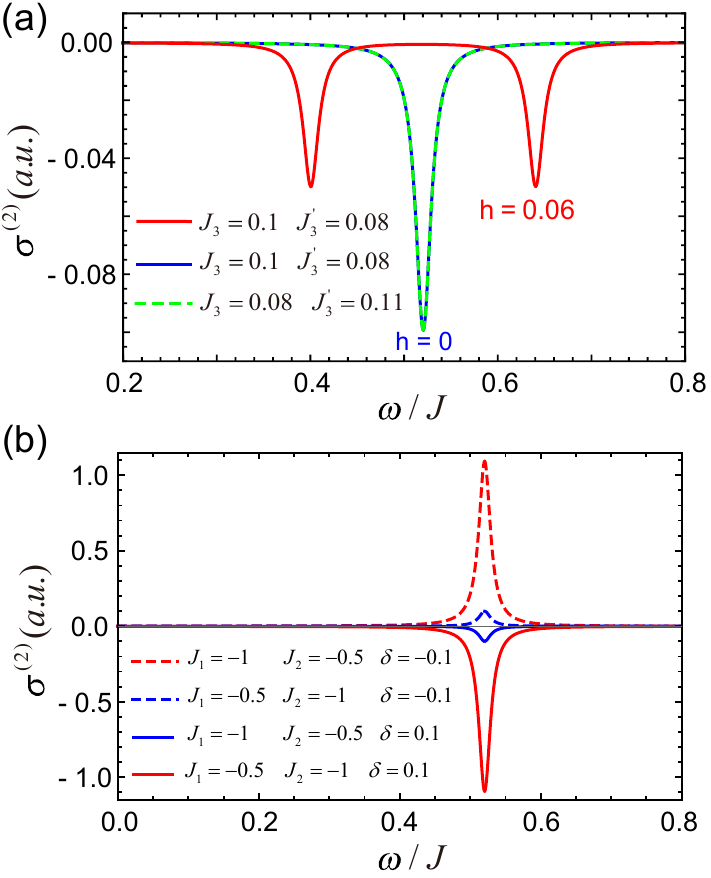}
  \caption{
    Nonlinear spin current conductivity $\sigma_{xx}^z(0;\omega,-\omega)$ for the 1D model in Eq.~\eqref{eq:1D_Hamiotlnian_initial}.  Frequency dependence of the spin current conductivity for (a) $J_1= -1$, $J_2 = -0.5$, $\delta= 0.1$, $D_z = 0.005$, and $\eta = 0.01$, and for (b) $J_3 = 0.1$, $J_3'=0.08$, and $h = 0$.
  }
  \label{fig:1D_conductivity}
\end{figure}

The frequency dependence of $\sigma_{zxx}^{(2)}$ for this model is shown in Fig.~\ref{fig:1D_conductivity}.
As shown in the figure, the resonance peaks appear in $\sigma_{zxx}^{(2)}$ at the frequency corresponding to the frequency of the magnons at $k=0$, indicating that a dc spin current is generated through a nonlinear optical process.
We note that, unlike the spin pump, this phenomenon occurs in a bulk magnetic insulator, and the spin current flows along the direction of magnetization.
When $h=0$, $\sigma_{zxx}^{(2)}$ shows a peak at $\omega\sim0.52$ reflecting the two degenerate magnon bands [Fig.~\ref{fig:1D_conductivity}(a)].
The peak splits into two under a static magnetic field, reflecting the lifting of the degeneracy of magnon bands in Fig.~\ref{fig:method_model}(e). The results indicate that, in addition to the spin current by a two-magnon process~\cite{Ishizuka2019b}, the spin current occurs by a one-magnon process.

In Fig.~\ref{fig:1D_conductivity}(a), the dashed green line shows the result for different $J_3$ and $J_3'$, which is exactly the same as the blue curve; $\sigma_{zxx}^{(2)}$ does not depend on $J_3$ and $J_3'$.
To better understand this behavior, let us look into the form of $\tilde\beta_{\bm 0}^x$.
For the model in Eq.~\eqref{eq:1D_Hamiotlnian_initial}, we find that $\tilde\beta_{\bm 0}^x$ has a special form,
\begin{align}\label{eq:1D_beta_form}
  \tilde\beta_{\bm 0}^x = ([\tilde{\beta}_{\bm 0}^\nu]_1, [\tilde{\beta}_{\bm 0}^\nu]_2, 0, 0, [\tilde{\beta}_{\bm 0}^\nu]_1, [\tilde{\beta}_{\bm 0}^\nu]_2, 0 ,0).
\end{align}
For Eq.~\eqref{eq:1D_beta_form}, the formula for $\sigma^{(2)}$ without the static magnetic field becomes:
\begin{align}\label{eq:1D_general_conductivity}
  [\sigma^{(2)}]_{xx}^z = \frac{\mathcal{- B}}{2\pi \hbar^2 V} \frac{\omega_{1,\bm 0} (6i\eta - 4\omega)}{(4\omega_{1,\bm 0}^2 + \eta^2)(\omega_{1,\bm 0}^2 + (\eta + i\omega)^2)},
\end{align}
where $\mathcal{B}$ reads
\begin{align}
  \mathcal{B} = & [\tilde{\beta}_{\bm 0}^x]_1 [\tilde{\beta}_{\bm 0}^x]_2\left([\tilde{\mathcal{J}}_{z, \bm 0}]_{25}+[\tilde{\mathcal{J}}_{z, \bm 0}]_{16} \right) \nonumber \\
                & + [\tilde{\beta}_1^x]^2 [\tilde{\mathcal{J}}_{z, \bm 0}]_{15} + [\tilde{\beta}_2^x]^2 [\tilde{\mathcal{J}}_{z, \bm 0}]_{26}.
\end{align}
In the 1D model, the first eigenfrequency $\omega_{1, \bm k}$ is independent of $J_3$ and $J_3'$. In addition, $\mathcal{B}$ does not change by changing $J_3$ and $J_3'$. Hence, the conductivity is independent of $J_3$ and $J_3'$.

Figure~\ref{fig:1D_conductivity}(b) shows the frequency dependence of $\sigma_{zxx}^{(2)}$ for $\delta<0$. The results indicate that, by switching $J_1$ and $J_5$ and changing $\delta \to -\delta$, the conductivity changes sign. This observation aligns with the spatial inversion operation about a site center, which is equivalent to the above transformation.
As the spin current changes sign by the inversion operation and the magnetic field does not, the nonlinear conductivity defined in Eq.~\eqref{eq:general_spincurrent_defination} should satisfy
\begin{align}\label{eq:1D_symmetry_final}
  \sigma^{(2)} (\delta^\ast) = -\sigma^{(2)} (-\delta^\ast),
\end{align}
where $\delta^\ast$ is an abstract parameter representing the non-centrosymmetry, i.e., the system changes $\delta^*\to-\delta^*$ by the inversion operation.
Hence, $\sigma_{zxx}^{(2)}$ changes sign by swapping $J_1$ and $J_5$, and changing $\delta^\ast \to -\delta^\ast$.

\section{Chromium oxide}\label{sec:3D}

\begin{figure}
  \centering
  \includegraphics[width=\linewidth]{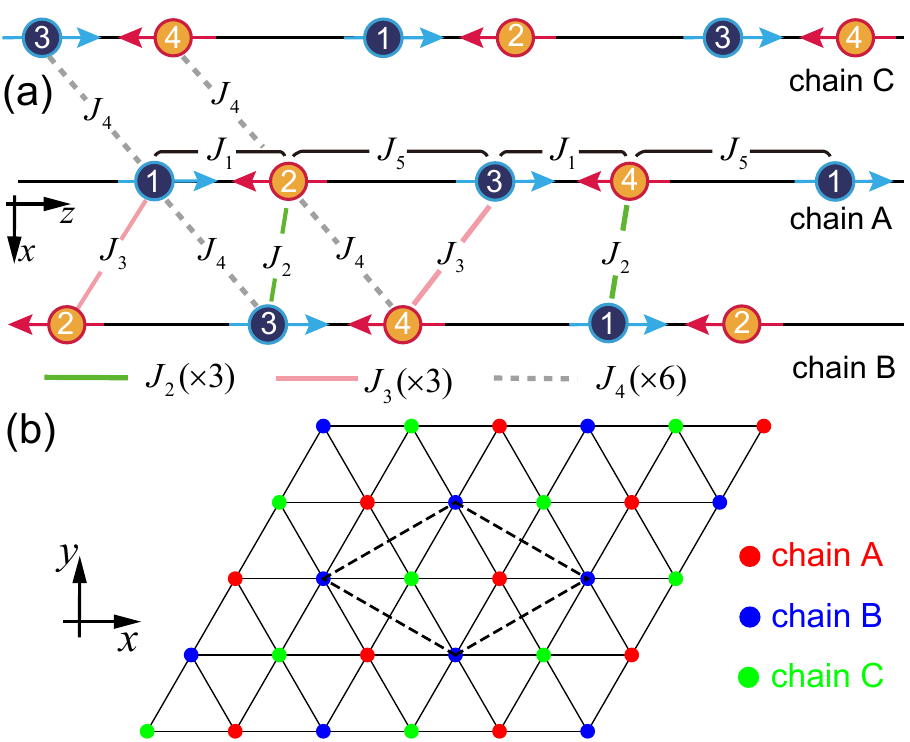}
  \caption{The crystal structure of Cr$_2$O$_3$.
    (a) A schematic of the crystal structure and exchange interactions $J_1$-$J_5$ viewed from $y$ axis.
    Numbers in the parentheses indicate the number of bonds.
    (b) The lattice structure viewed from the $z$ axis. The $A$, $B$, and $C$ chains correspond to the respective chains in (a).
    The dashed rhombus shows the unit cell of Cr$_2$O$_3$.
  }
  \label{fig:3D_model_diagram}
\end{figure}

In this section, we consider chromium sesquioxide (Cr$_2$O$_3$) as an example.
Cr$_2$O$_3$ is an antiferromagnetic insulator with a trigonal corundum crystal structure (space group $R\bar{3}c$)~\cite{Schober1995a,Hill2010a} which is known for magnetoelectric (ME) effect~\cite{Dyaloshinskii1960a,Astrov1960a,Rado1961a,Iyama2013a}.
There are four Cr atoms in a unit cell aligned along the $z$ axis; the ground state is an antiferromagnetic order with up-down-up-down configuration shown in Fig.~\ref{fig:3D_model_diagram}(a)~\cite{Brockhouse1953a,Mcguire1956a,Corliss1965a}.
This unique arrangement breaks the inversion symmetry, the magneto-electric effects, and the rectification of spin current through a nonlinear response~\cite{Ishizuka2019b}.

\begin{figure}
  \centering
  \includegraphics[width=\linewidth]{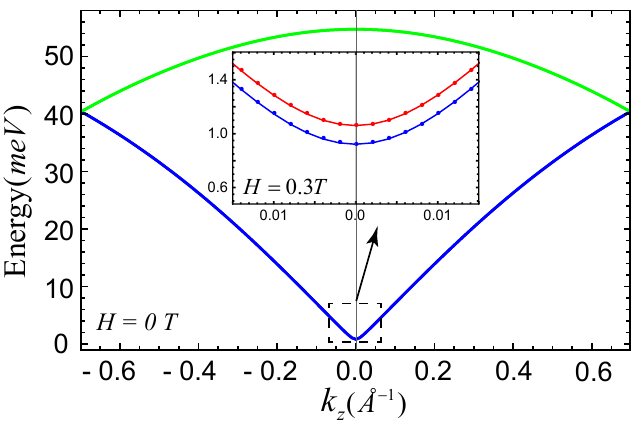}
  \caption{
    Magnon dispersion for the model in Eq.~\eqref{eq:3dmodel} with $h=0$ and $h=0.3$ T (inset).
    The results are for $J_1 = -7.53$ meV, $J_2 = -3.41$ meV, $J_3 =-0.08$ meV, $J_4 = 0.02$ meV, $J_5 = -0.19$ meV~\cite{Samuelsen1970a}, and $D_z = 0.0015$ meV~\cite{Foner1963a, Mu2019a}.
  }
  \label{fig:3D_magnon_dispersion}
\end{figure}

The effective model for Cr$_2$O$_3$ was investigated by Samuelsen {\it et al}.~\cite{Samuelsen1970a}, which is shown in Fig.~\ref{fig:3D_model_diagram}. The projected view from the $z$ axis is shown in Fig.~\ref{fig:3D_model_diagram}(b).
The model consists of five exchange interactions.
Along the chain, $J_1$ and $J_5$ are the nearest-neighbor (NN) interactions in the main chain, each with a coordination number of 1.
In addition, $J_2$ and $J_3$ are the interactions between neighboring chains, each with a coordination number of 3.
In the last, $J_4$ is the exchange interaction between the spins on different chains with a coordination number of 6.
The Hamiltonian reads
\begin{align}
  \mathcal{H}^{3D} = \mathcal{H}_{1}^{3D} + \mathcal{H}_{2}^{3D} + \mathcal{H}_D + \mathcal{H}_h,\label{eq:3dmodel}
\end{align}
where the form of $\mathcal{H}_D$ and $\mathcal{H}_h$ are same as the one in Eq.~\eqref{eq:1D_Hamiotlnian_initial}, and the other two terms,
\begin{align}
  \mathcal{H}_{1}^{3D} = & - \sum\limits_{n} \sum\limits_j J_1{{\bm S}_{n,j,1}} \cdot {{\bm S}_{n,j,2}} + J_1{{\bm S}_{n,j,3}} \cdot {{\bm S}_{n,j,4}}\nonumber \\
                         & \qquad\qquad+ J_5{{\bm S}_{n,j,2}} \cdot {{\bm S}_{n,j,3}} + J_5{{\bm S}_{n,j,4}} \cdot {{\bm S}_{n,j+1,1}}, \nonumber               \\
  \mathcal{H}_{2}^{3D} = & - \sum\limits_{\langle n,m \rangle
  } \sum\limits_j J_3 ({{\bm S}_{n,j,1}} \cdot {{\bm S}_{m,j,2}} + {{\bm S}_{n,j,3}} \cdot {{\bm S}_{m,j,4}})\nonumber                                          \\
                         & \qquad\qquad+ J_2({{\bm S}_{n,j,2}} \cdot {{\bm S}_{m,j,3}} + {{\bm S}_{n,j,4}} \cdot {{\bm S}_{m,j+1,1}}) \nonumber                 \\
                         & \qquad\qquad+ J_4 ({{\bm S}_{n,j,1}} \cdot {{\bm S}_{m,j,3}} + {{\bm S}_{n,j,2}} \cdot {{\bm S}_{m,j,4}}),
\end{align}
are the interactions between different Cr ions in the same chain and neighboring chains, respectively.
Here, $\bm{S}_{n,j,u}$ is the operator for spins on the $u (=1,2,3,4)$ sublattice in the $j$th unit cell in the $n$th chain.
The results of the magnon band are plotted in Fig.~\ref{fig:3D_magnon_dispersion}.
Similarly to the one-dimensional model, the magnon bands are doubly degenerate at $h = 0$, whereas a finite magnetic field $h$ can lift the degeneracy, as shown in the inset of Fig.~\ref{fig:3D_magnon_dispersion}.

\begin{figure*}[t]
  \centering
  \includegraphics[width=\linewidth]{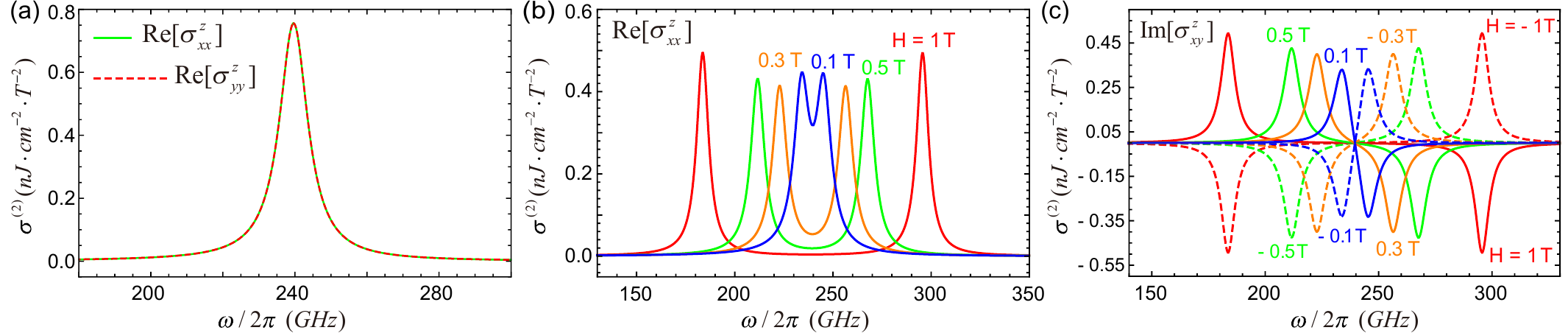}
  \caption{
    Nonlinear spin current conductivity $\sigma^z(0;\omega,-\omega)$ with Gilbert damping $\alpha = 0.02$, $J_1 = - 7.53$ meV, $J_2 = -3.41$ meV, $J_3 = -0.08$ meV, $J_4 = 0.02$ meV, $J_5 = -2.2$ meV,  and $D_z = 0.0015$ meV. Frequency dependence of (a) $[\sigma^{(2)}]_{xx}^z$ and $[\sigma^{(2)}]_{yy}^z$ at $h=0$, (b) $[\sigma^{(2)}]_{xx}^z$, and (c) $[\sigma^{(2)}]_{xy}^z$. The dashed curves in (c) are the results for negative magnetic fields. }
  \label{fig:3D_conductivity_combine}
\end{figure*}

The frequency dependence of $[\sigma^{(2)}]^z_{xx}(0; \omega, -\omega)$ is shown in Fig.~\ref{fig:3D_conductivity_combine}. Here, the magnon relaxation rate is taken to be $\eta = \alpha \omega$~\cite{Abrikosov1963a}, where $\alpha = 0.02$ is the damping constant for Cr$_2$O$_3$ known in experiment~\cite{Nguyen2020a}.
The results for $\left[\sigma^{(2)}\right]_{yy}^z$ is the same as $[\sigma^{(2)}]^z_{xx}$.
Similar to the one-dimensional case, a resonance peak appears in Fig.~\ref{fig:3D_conductivity_combine}(a) at a frequency corresponding to the frequency of magnons at $\bm k = \bm 0$. The conductivity at the resonance peak in Fig.~\ref{fig:3D_conductivity_combine}(a) is $\sigma^{(2)} \sim 10^{-9}$ J$\cdot$cm$^{-2}$$\cdot$T$^{-2}$, which is similar to the value in a previous study~\cite{Ishizuka2022a}. Figure~\ref{fig:3D_conductivity_combine}(b) shows the magnetic field dependence of the real part of conductivity $\left[\sigma^{(2)}\right]_{xx}^z$. Under a magnetic field, the resonance peak splits into two, reflecting the lifting of the degeneracy of the magnon bands [Fig.~\ref{fig:3D_magnon_dispersion}]. By increasing the external static magnetic field, the lower band eventually reaches zero energy, destabilizing the antiferromagnetic order; the field-induced phase transition occurs at $H \sim 3.5-4.0$ T for the parameters we used.

  Next, we look into the spin photocurrent induced by a circularly polarized light~\cite{Proskurin2018a}.
  Figure~\ref{fig:3D_conductivity_combine}(c) shows the magnetic field dependence of the imaginary part of conductivity $\left[\sigma^{(2)}\right]_{xy}^z$, which corresponds to the spin current induced by circularly polarized light.
  At $h=0$, the spin photocurrent vanishes in contrast to the case of linearly-polarized light, and a finite spin current occurs under the magnetic field.
  In addition, the direction of the spin current inverts by reversing the magnetic field as shown by the dashed curves in Fig.~\ref{fig:3D_conductivity_combine}(c).
  The results show that the resonance frequency and the direction of spin photocurrent are controllable by using an external magnetic field.

  \section{Discussion and Conclusion}\label{sec:discussion}

  In this work, we investigated the magnon spin photocurrent in Cr$_2$O$_3$ based on a four-sublattice spin model.
  Using the nonlinear response theory, we found that the conductivity exhibits a single response peak when there are
  different eigenfrequencies. Notably, both the position and height are independent of the NNN interaction in the one-dimensional effective model for Cr$_2$O$_3$.
  Additionally, we extended our study to the three-dimensional model for Cr$_2$O$_3$. Symmetry analysis revealed that the conductivity $\left[\sigma^{(2)}\right]^z(0;\omega,-\omega)$, under linearly and circularly polarized light, exists only along the $z$ direction.
  Furthermore, we study the conductivity for both linearly and circularly polarized light, and explore its tunability via an external magnetic field.

  Lastly, we discuss the magnitude of the induced spin current.
  Assuming a linearly-polarized electromagnetic wave of $H = 0.1 \sim 1$ mT amplitude, we find $\sigma^{(2)} \sim 10^{-9}$ J$\cdot$cm$^{-2}\cdot$T$^{-2}$ for the three-dimensional model in Sec.~\ref{sec:3D}.
  This value is comparable or larger than that in a previous research~\cite{Ishizuka2022a}.
  We further demonstrated that circularly polarized light also generates a spin photocurrent in the presence of an external magnetic field, giving a similar order of magnitude.
  The results imply that the spin photocurrent induced in Cr$_2$O$_3$ is likely to be observable in the experiment.
  Cr$_2$O$_3$ is stable in the air and has a high Neel temperature (307 K)~\cite{Kittel2018a}, facilitating experimental verification.

  \acknowledgements

  We are grateful to H. Murata and H. Yoshida for fruitful discussions.
  This work is supported by JSPS KAKENHI (Grant Numbers JP19K14649, JP23K03275, and JP25H00841) and JST PRESTO (Grant No. JPMJPR2452). Z.C.G acknowledges financial support from JST SPRING under Grant No. JPMJSP2180.

  \appendix

  \section{Spin wave Hamiltonian for collinear magnets}\label{sec:collinear}

  In this section, we summarize some of the basic properties of the magnon Hamiltonian for a collinear magnetic order that are relevant to the current study.
  To this end, we consider a general spin model with $n_{uc}$ sites in the unit cell,
  \begin{align}
    H=-\frac12\sum_{ia,jb}J_{ia,jb}\bm S_{ia}\cdot\bm S_{jb}-\sum_{ia}D_{a}(S_{ia}^z)^2,\label{eq:Hgen}
  \end{align}
  where $\bm S_{ia}$ is the spin on the $a$th sublattice in the $i$th unit cell.
  We further assume that the ground state is a collinear magnetic order with spins pointing along the $z$ axis; the direction of the spin is $n_a^z=\pm1$.

  Within the linear spin-wave approximation, the magnon Hamiltonian reads
  \begin{align}
    H=               & \frac12\sum_{\bm k}\psi_{\bm k}^\dagger\check H(\bm k)\psi_{\bm k}+\text{const.},\label{eq:Hmagnon} \\
    \check H(\bm k)= & \left(\begin{array}{cc}H(\bm k)&H'(\bm k)\\ H'(\bm k)&H(\bm k)\\\end{array}\right),
  \end{align}
  where $H(\bm k)$ and $H'(\bm k)$ being $n_{uc}\times n_{uc}$ matrices whose elements are
  \begin{align}
    H_{ab}(\bm k)=  & -\frac{(1+n_a^zn_b^z)\sqrt{S_aS_b}}2J_{ab}(\bm k)\nonumber                          \\
                    & +\delta_{ab}\left[2D_aS_a+\sum_{b}J_{a,b}(\bm k=\bm0)S_bn_a^zn_b^z\right],\nonumber \\
    H'_{ab}(\bm k)= & -\frac{(1-n_a^zn_b^z)\sqrt{S_aS_b}}2J_{ab}(\bm k),\nonumber
  \end{align}
  and $\psi_{\bm k}=\left(b_{1\bm k},\dots,b_{n_{uc}\bm k},b_{1,-\bm k}^\dagger,\dots,b_{n_{uc},-\bm k}^\dagger\right)^t$ is the vector of magnon annihilation (creation) $b_{i\bm k}$ ($b_{i\bm k}^\dagger$) operators with momentum $\pm\bm k$, and $J_{ab}(\bm k)=\sum_{i}J_{a,b}(\bm r_i)e^{-i\bm k\cdot(\bm r_i+\bm\delta_a-\bm\delta_b)}$.

  The Hamiltonian in Eq.~\eqref{eq:Hmagnon} can be diagonalized using the eigenstate vector $|u_{n\bm k}\rangle$ satisfying~\cite{Colpa1978a}
  \begin{align}
    \tau^z\check H(\bm k)|u_{n\bm k}\rangle=\gamma_n\omega_{n\bm k}|u_{n\bm k}\rangle,\label{eq:evalH}
  \end{align}
  where $\omega_{n\bm k}=\omega_{n+n_{uc},\bm k}$ and $\tau^z={\rm diag}(1,-1)\otimes1_{n_{uc}}$ with $1_{n_{uc}}$ being the $n_{uc}\times n_{uc}$ identity matrix.
  From here on, we consider normalized eigenstates satisfying $\langle u_{n\bm k}|\tau^z|u_{m\bm k}\rangle=\gamma_n\delta_{nm}$, and $|u_{n+n_{uc},-\bm k}\rangle=\tau^x|u_{n,\bm k}^\ast\rangle$.
  Using the eigenstates, the annihilation and creation operators for the magnons in the diagonal basis read $\phi_{n\bm k}=\gamma_n\langle u_{n\bm k}|\tau^z\psi_{\bm k}$, where $n\le n_{uc}$ are the annihilation operator for $\omega_{n\bm k}$ mode and $n>n_{uc}$ are the creation operators for $\omega_{n-n_{uc},\bm k}$ mode.

  In addition, for the collinear magnets in Eq.~\eqref{eq:Hmagnon}, the following relations hold:
  \begin{align}
    [\tau^x,\check H(\bm k)]=0,\quad \check H(-\bm k)=\check H^t(\bm k).
  \end{align}
  The second equality corresponds to the effective time-reversal symmetry assumed in a recent work~\cite{Fujiwara2023a}.
  However, here, it is a property shown to be satisfied in any models of the form in Eq.~\eqref{eq:Hgen} with a collinear magnetic order.
  In such a case, $|u_{n\bm k}\rangle$ satisfies $\omega_{n\bm k}=\omega_{n,-\bm k}$ and
  \begin{align}
    |u_{n,-\bm k}\rangle=|u_{n,\bm k}^\ast\rangle,\quad |u_{n+n_{uc},\bm k}\rangle=\tau^x|u_{n,\bm k}\rangle.
  \end{align}

  Lastly, we briefly discuss the properties of the matrix $\check n^z=1_2\otimes{\rm diag}(n^z_1,\cdots,n^z_{n_{uc}})$, for the sake of later convenience. This matrix is related to the net magnetization,
  \begin{align}
    \sum_{ia}S_{ia}^z=NS-\frac12\sum_{\bm k}\psi_{n\bm k}^\dagger \check n^z\psi_{n\bm k}.
  \end{align}
  Considering that $n_a^z=\pm1$, for the magnon Hamiltonians in Eq.~\eqref{eq:Hmagnon}, one can show by explicit calculation that
  \begin{align}
    [\tau^z\check n,\tau^z\check H(\bm k)]=[\check n\tau^z,\check H(\bm k)\tau^z]=0.
  \end{align}
  The commutation relation implies that there exists a set of $|u_{n\bm k}\rangle$ that simultaneously diagonalize $\tau^z\check n$ and $\tau^z\check H(\bm k)$.
  To be concrete, let us consider a matrix whose elements are given by $(T_{\bm k})_{ab}=|u_{b\bm k}\rangle_a$, where $|u_{b\bm k}\rangle_a$ is the $a$th element of $|u_{b\bm k}\rangle$ that simultaneously satisfies Eq.~\eqref{eq:evalH} and
  \begin{align}
    \tau^z\check n|u_{n\bm k}\rangle=\gamma_n\tilde n_{n\bm k}|u_{n\bm k}\rangle.\label{eq:evalN}
  \end{align}
  We also note that $T^\dagger_{\bm k}\tau^zT_{\bm k}=\tau^z$~\cite{Colpa1978a}.
  By using $T_{\bm k}$, one can diagonalize
  \begin{align}\label{eq:Appendix_Spinwave_bogoliubov}
    T_{\bm k}^\dagger\check H(\bm k)T_{\bm k} =E_{\bm k},\quad T_{\bm k}^\dagger\check n^zT_{\bm k} =\tilde n_{\bm k},
  \end{align}
  where $E_{\bm k}=1_2\otimes{\rm diag}(\omega_{1\bm k},\cdots,\omega_{n_{uc}\bm k})$.

  \section{Derivation of the Berry phase formula}\label{sec:Berry}

  Within the linear spin wave approximation, the spin current operator in Eq.~\eqref{eq:Jspin-gen} for collinear magnets in Eq.~\eqref{eq:Jspin-gen} reads,
  \begin{align}
    \mathcal J^{\mu}=\frac12\sum_{\bm k}\psi^\dagger(\bm k)\left[\check n^z\tau_z\partial_{k_\mu}\check H(\bm k)+\partial_{k_\mu}\check H(\bm k)\tau_z\check n^z\right]\psi(\bm k).
  \end{align}
  For the sake of conciseness, we use this notation for the spin current in this section.
  Using this spin current operator, the spin current conductivity for a general perturbation $\beta^\nu_{l\bm k}$ reads
  \begin{align}\label{eq:appendix_berry_conductivity}
     & \sigma^{(2)}_{\lambda\mu\nu}(0; \omega, - \omega) =- \frac{1}{2\pi V} \nonumber \\
     & \sum_{l,m=1}^{2n_{uc}} \sum_{\bm{k}}
    \frac{\gamma_l \gamma_m}
    {\hbar\omega - \gamma_l \hbar\omega_{\bm{k}} - i\eta}
    \frac{\tilde{\beta}^{\mu}_{\bm{k}} [\tilde{\mathcal{J}}_{\lambda,\bm k}]_{lm} (\tilde{\beta}^\nu_{m\bm{k}})^\ast\hat B_{\bm k}^\mu(\hat B_{\bm k}^\nu)^\ast}
    {- \gamma_l \hbar\omega_{l\bm{k}} + \gamma_m \hbar\omega_{m\bm{k}} - i\eta}.
  \end{align}
  Here,
  \begin{align}\label{eq:appendix_berry_sc}
    [\tilde{\mathcal J}_{\lambda\bm k}^\mu]_{lm} = \frac12 \sum_{\bm k} \bra{u_{l\bm k}} \left(\check n^z\tau_z\partial_{k_\mu}\check{H}_{\bm k}+\partial_{k_\mu}\check {H}_{\bm k}\tau_z\check n^z\right) \ket{u_{m\bm k}},
  \end{align}
  is the spin current operator in the eigenstate basis, and
$\tilde{\beta}_{l \bm k}^\mu$ is the component of the perturbation in the eigenstate basis, $\mathcal{H}'=-\sum_{l,\bm k,\mu}B_\mu(t)\hat B_{\bm k}^\mu\beta_{l\bm k}^\mu \psi_{l\bm k}$.
  It is related to $\beta_{l \bm k}^\mu$ by \begin{align}\label{eq:appendix_berry_beta}
    \tilde{\beta}_{l \bm k}^\mu = \sum_a \left[\beta_{l \bm k}^{\mu}\right]_a \ket{u_{l \bm k}}_a.
  \end{align}
  The $\hat B_{\bm k}^\mu$ is a function reflecting the real-space structure of the perturbation; it is $\hat B_{\bm k}^\mu=\delta_{\bm0,\bm k}$ for a spatially-uniform perturbation.

  Substituting Eq.~\eqref{eq:appendix_berry_sc} into Eq.~\eqref{eq:appendix_berry_conductivity}, the conductivity reads
  \begin{align}
    \sigma^{(2)}_{\lambda\mu\nu}(0; \omega, - \omega)   & =\sigma^{(2;1)}_{\lambda\mu\nu}(0; \omega, - \omega)+\sigma^{(2;2)}_{\lambda\mu\nu}(0; \omega, - \omega),                                                                                                                                      \\
    \sigma^{(2;1)}_{\lambda\mu\nu}(0; \omega, - \omega) & =\nonumber                                                                                                                                                                                                                                     \\
    - \frac{i}{2 \pi V \eta}\sum_{\substack{
        l,\,\bm{k}
    }}                                                  & \frac{ \tilde{\beta}^{\mu}_{l\bm{k}} \tilde{n}_{l \bm k} \gamma_l \partial_{k_\lambda} \omega_{l\bm k}(\tilde{\beta}^{\nu}_{l\bm{k}})^\ast }{\hbar\omega - \gamma_l \hbar\omega_{l\bm{k}} - i\eta}\hat B_{\bm k}^\mu(\hat B_{\bm k}^\nu)^\ast,
  \end{align}
  \begin{align}
    \sigma^{(2;2)}_{\lambda\mu\nu} & (0; \omega, - \omega)=\frac{i}{2\pi V} \sum_{\substack{
    l \neq m; \, \bm{k}                                                                                                              \\a, b}}
    \frac{\gamma_l \gamma_m}{\hbar\omega - \gamma_l \hbar\omega_{l\bm{k}} - i\eta}\times\nonumber                                    \\
    g^\mu_a                        & \ket{u_{l\bm{k}}}_a (\tilde{n}_l \gamma_l+\gamma_m\tilde{n}_m ) \mathcal{A}_{lm}^\lambda(\bm k)
    \bra{u_{m\bm{k}}}_b (g^\nu_b)^*\hat B_{\bm k}^\mu(\hat B_{\bm k}^\nu)^\ast,\label{eq:appendix_berry_nmsigma}
  \end{align}
  where $\mathcal{A}_{lm}^\lambda(\bm k)$ is the Berry connection defined in Sec.~\ref{sec:model}.
  Here, we note that $\bra{\partial_{k_\lambda} u_{l \bm k}}=\sum_m i \gamma_m \mathcal{A}_{lm}^\lambda(\bm k) \bra{u_{m \bm k}}$.
  Using this equality, Eq.~\eqref{eq:appendix_berry_nmsigma} can be simplified as
  \begin{align}
     & \sigma^{(2;2)}_{\lambda\mu\mu}(0; \omega, - \omega) =\nonumber \\
     & - \frac{i}{2\pi V} \sum_{
      l,\bm k,\mu}
    \frac{\tilde{n}_l  \left[ \partial_{k_\lambda}\phi^\mu_{l \bm k} + \gamma_l \mathcal{A}_{ll}^\mu(\bm k) \right]}{\hbar\omega - \gamma_l \hbar\omega_{l\bm{k}} - i\eta}  \left|\tilde{\beta}_{l \bm k}^\mu\right|^2|\hat B_{\bm k}^\mu|^2,
  \end{align}
  where $\phi^\mu_{l \bm k}$ is defined by $\tilde{\beta}_{l \bm k}^\mu=|\tilde{\beta}_{l \bm k}^\mu|e^{i\phi^\mu_{l \bm k}}$.
  For the uniform external field, $\hat B_{\bm k}^\mu=\delta_{\bm0,\bm k}$.
  Hence, the conductivity reads
  \begin{align}
     & \sigma^{(2)}_{\lambda\mu\mu}(0;\omega,-\omega) =\nonumber \\
     & - \frac{i}{2\pi V} \sum_{
      l}
    \frac{\tilde{n}_l  \left[ \partial_{k_\lambda}\phi^\mu_{l \bm k} + \gamma_l \mathcal{A}_{ll}^\mu(\bm k) \right]_{\bm k=\bm0}}{\hbar\omega - \gamma_l \hbar\omega_{l\bm0} - i\eta}  \left|\tilde{\beta}_{l\bm0}^\mu\right|^2,
  \end{align}
  which is Eq.~\eqref{eq:shift-current}.
  Here, we assumed $\partial_{k_\lambda} \omega_{l,\bm k} =0$ for $\bm k=0$.

This equation resembles a similar equation for the two-magnon process, which is related to the difference of the Berry connection~\cite{Fujiwara2023a}.
However, unlike the case of the two-magnon process, here the spin current conductivity is related to the Berry connection of the excited magnon.Note that a similar formula was recently given for the Peltier effect of phonons~\cite{Ishizuka2024a}.
The coincidence appears to show that the relation between the response to linearly polarized light and shift-current-like physics is universal regardless of the type of quasi-particles.
It shows that the magnon photocurrent resulting from the single-magnon process is also related to the shift of position, similar to the shift current of electrons~\cite{vBaltz1981a}.

\bibliography{ref}

\end{document}